**Assessing Public Perception of Car Automation in Iran: Acceptance and Willingness to Pay for Adaptive Cruise Control**

**Sina Sahebi**
Assistant Professor
Faculty of Civil, Water and Environmental Engineering
Shahid Beheshti University, Tehran, Iran
Email: s_sahebi@sbu.ac.ir

**Sahand Heshami**
Master student of transportation engineering
Faculty of Civil, Water and Environmental Engineering
Shahid Beheshti University, Tehran, Iran
Email: s.heshami@mail.sbu.ac.ir

**Mohammad Khojastehpour**
Graduate Research Assistant
Department of Civil & Environmental Engineering
The University of Tennessee, TN 37996, USA
Email: Mkhojast@vols.utk.edu

**Ali Rahimi**
Master student of transportation engineering
Faculty of Civil, Water and Environmental Engineering
Shahid Beheshti University, Tehran, Iran
Email: a.rahimi@mail.sbu.ac.ir

**Mahyar Mollajani**
Master student of transportation engineering
Faculty of Civil, Water and Environmental Engineering
Shahid Beheshti University, Tehran, Iran
Email: m.mollajani@mail.sbu.ac.ir

Word count: 4978 words text + 8 tables x 250 words (each)+ 5 figures x 0 words= 6978 words

**ABSTRACT**

Adaptive cruise control (ACC) is a technology that can reduce fuel consumption and air pollution in the automotive industry. However, its availability in Iran is low compared to industrialized countries. This study examines the acceptance and willingness to pay (WTP) for ACC among Iranian drivers. Data from an online survey of 453 respondents were analyzed using the Technology Acceptance Model (TAM) and an ordered logit model. The results show that perceived ease of use and perceived usefulness affect attitudes toward using ACC, which in turn influence behavioral intentions. The logit model also shows that drivers who find ACC easy and useful, who have higher vehicle prices, and who are women with cruise control (CC) experience are more likely to pay for ACC. To increase the adoption of ACC in Iran, it is suggested to target early adopters, especially women and capitalists, who can influence others with their positive feedback. The benefits of ACC for traffic safety and environmental sustainability should also be emphasized.

**INTRODUCTION**

Adaptive cruise control (ACC) is a driver-assist feature that helps with the task of longitudinal control of a vehicle during motorway driving. The system controls the accelerator, engine powertrain, and vehicle brakes to maintain a desired time-gap to the vehicle ahead (1), (2). ACC allows drivers to maintain a fixed speed and sustain a set distance from the car ahead of them, and it can increase road safety by keeping adequate spacing between vehicles and preventing accidents that result from an obstructed view (3), (4). The integration of advanced technologies in the transportation sector has been a key focus for policymakers aiming to reduce fuel consumption and carbon dioxide ($CO_2$) emissions. Cruise Control (CC) has traditionally been used to control traffic and offers benefits such as maintaining vehicle speed, reducing fuel consumption, and enhancing driver comfort during long-distance journeys (5). However, to address environmental concerns and improve energy efficiency, the concept of eco-driving has gained widespread attention, considering traffic flow and road information (6,7). A study by Liu et al. in 2021 highlighted the direct effect of ACC and CC in reducing fuel consumption on freeways (8). The transportation sector, predominantly powered by internal combustion engines, contributes to the emission of harmful substances including carbon monoxide (CO), nitrogen oxides (NOx), particulates (PM), and hydrocarbons (HC) (9). Additionally, the combustion process results in significant carbon dioxide ($CO_2$) emissions, contributing to global warming (10). In Europe, the transportation system alone accounts for over a quarter of total $CO_2$ emissions (11). To meet emission standards, advanced technologies have been introduced to enhance engine and drive system efficiency (12,13). One such advancement is ACC, an improvement over CC that helps vehicles maintain a safe distance from other cars by controlling throttle and braking. ACC operates through radar or laser-based systems, detecting moving objects and adjusting vehicle speed to match that of the front vehicle. ACC falls under level 1 of driver assistance systems according to the Society of Automotive Engineers (SAE) classification (14). By utilizing radar or lidar technology, ACC accurately determines distances even in adverse weather conditions. When a slower-moving vehicle is detected ahead, ACC reduces torque and activates the braking system to maintain the desired distance (15,16). ACC represents an important step in enhancing driver safety, comfort, reducing fuel consumptions, and overall driving experience.

Iran as a developing country, is facing significant challenges related to traffic safety and air pollution. Iran is located in the region where traffic-related death rates are the second highest in the world (17) The country has a serious problem with high traffic levels due to several factors, including transportation strategies and sociocultural and economic features (18). Despite efforts to improve traffic safety, the number of deaths due to road traffic injuries (RTIs) has exceeded deaths from heart diseases in Iran (19). Air pollution is also a significant problem in Iran, particularly in major cities such as Tehran (20).

The utilization of ACC in Iranian vehicles has the potential to alleviate both traffic safety concerns and environmental issues in the country, as supported by its benefits. In contrast, awareness and knowledge about ACC among individuals in developing countries like Iran are limited. Despite the efforts of car manufacturers to incorporate advanced features in their products, many Iranians remain unfamiliar with technologies such as ACC. Nevertheless, ACC is an essential and inseparable component of modern automobiles, providing drivers with enhanced efficiency, safety, and ease of use. This emphasizes the importance of raising awareness and educating the public about the benefits of ACC in Iran.

Public acceptance is a key factor for the success of any new technology, including new automotive technologies. (21). Factors that influence public acceptance of new automotive technologies include perceived safety, perceived usefulness, perceived ease of use, perceived risk, perceived satisfaction, self-efficacy, technological maturity, technological standards, research funds, national or regional support policies, infrastructure facilities, and costs (22,23,24,25,26).

In recent times, there has been significant discourse and deliberation among scholars concerning the contentious matter of the impact of drivers' opinions towards ACC on their willingness to purchase it at a reasonable cost in developing countries. Moreover, the present paper seeks to fill the research gap by examining the public acceptance of and willingness to pay (WTP) for ACC systems in developing countries. The contribution of this paper is threefold. Firstly, it addresses the pressing concerns of safety and air pollution in developing countries, highlighting how the adoption of advanced automotive technologies, specifically ACC, can help mitigate these issues. By exploring the penetration rate improvement of ACC in a developing country, this study provides valuable insights into how to enhance the adoption and utilization of this technology for maximum impact. Secondly, this research recognizes ACC as a significant component of car intelligentization, which encompasses driver assistant and self-driving technologies. By measuring the acceptance of ACC usage, this study serves as a crucial criterion and tool for understanding the overall development of automobiles and the advancement of driver assistant technologies in developing countries. This assessment sheds light on the current state and potential growth of intelligent automotive systems in these regions. Lastly, the findings of this study have implications for producers, car manufacturers, and other relevant stakeholders in the automotive industry. Understanding the market potential and acceptance of ACC in a developing country can guide producers and manufacturers in their strategies to promote and integrate this technology effectively. The insights gained from this research contribute to the knowledge base on sustainable transportation solutions, enabling stakeholders to make informed decisions and advancements in driving safety, air pollution reduction, and overall automotive development within developing nations. This study may help to inform future research on the topic in developing countries and other countries with similar cultural and socio-economic contexts. Figure 1 indicates the study flowchart.

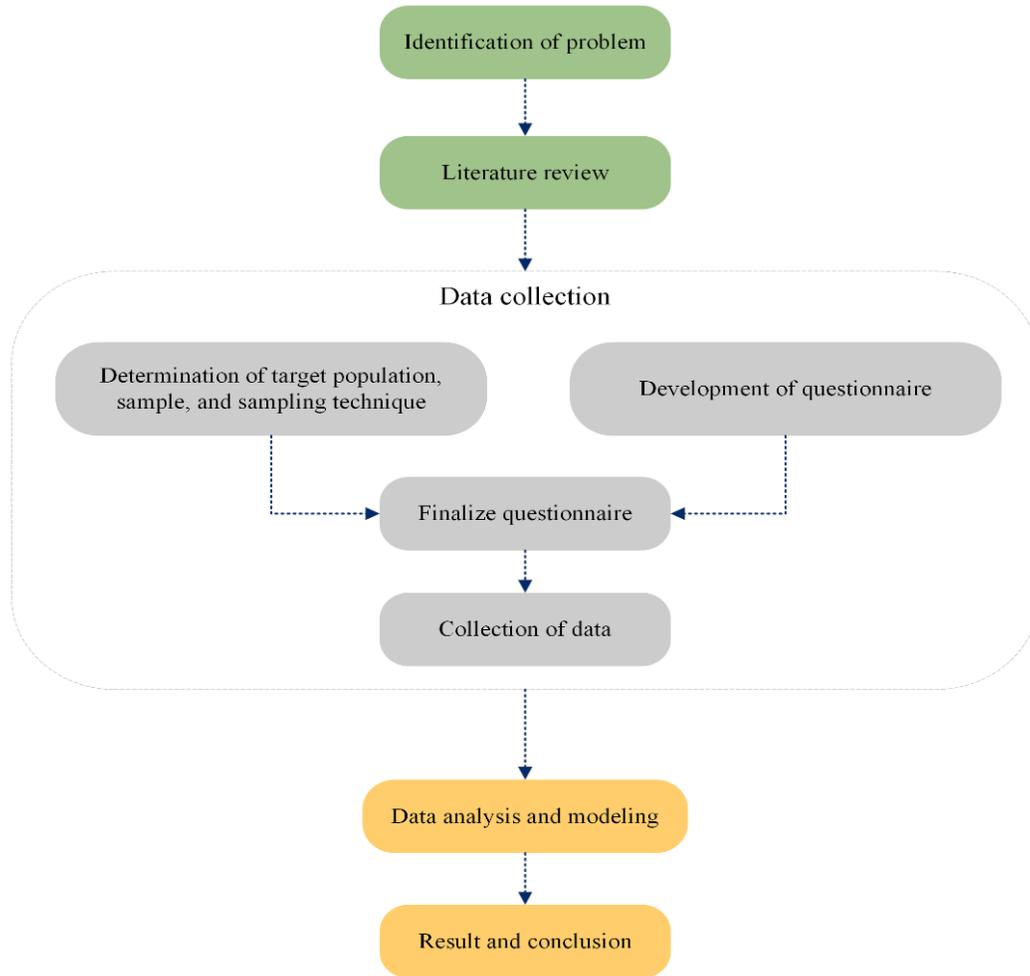

**Figure 1.** Study flowchart

**LITERATURE REVIEW**

ACC has been the subject of various studies, highlighting both its benefits and potential drawbacks. Abendroth conducted a study indicating that drivers using ACC on highways chose a mean speed of 119 km/h, compared to 129 km/h without ACC, resulting in reduced speed violations exceeding 20 km/h (27). Similar effects were observed in field operational tests (28,29) and a driving simulator study (30). However, contrasting findings have also been reported. Hoedemaeker et al. conducted a driving simulator study with 38 participants and found that drivers using ACC chose smaller headways and drove at higher speeds compared to manual driving (31). Buld and Krüger highlighted that drivers with ACC tended to maintain the same inadequate speed as the lead vehicle in tight turns and exhibited decreased performance in lane-keeping tasks (32). These studies provide insights into both the positive and negative effects of ACC on driving behavior and safety. A survey conducted by Öörni in 2016 revealed that in the European market, the adoption of ACC was relatively low, with 1.7% and 3.2% of cars sold in 2011 and 2013, respectively, being equipped with ACC. However, Germany stood out as the frontrunner with adoption rates of 3.7% and 6.8% during the same years (33). Similarly, a survey conducted among U.S. drivers showed that 14% of respondents had ACC installed in their vehicles (34). Moreover, the use of ACC has been found to negatively impact drivers' lane-keeping performance. While the reduction in workload associated with ACC may initially appear beneficial, it has been observed that situational awareness is diminished when using this technology (35,36,37). Situational awareness, a concept frequently studied in human factors research, refers to the ability to perceive and understand environmental cues and anticipate changes (38). Drivers who heavily rely on the ACC system may experience a decline in situational awareness, as their attention may be diverted from their surroundings (38). Furthermore, the theory of changeable attentional resources proposed by Young and Stanton suggests that the diminished situational awareness could be a result of depleted attentional resources (39).

In terms of performance, drivers using ACC have shown higher overall performance; however, their response times to potential safety issues, such as the need to brake, have been found to be longer (40). This highlights the potential trade-off between improved driving performance under normal conditions and the need for quick responses in critical situations, although field tests have shown that ACC can increase distances between leading cars and improve adherence to speed limits (27,28,29), there is also compelling evidence that drivers may struggle to maintain situational awareness, leading to delayed response times and potential distractions from driving-related tasks (30,31,32,41). Additionally, the decreased workload brought on by ACC may result in a depletion of attentional resources.

The Technology Acceptance Model (TAM), initially proposed by Davis in 1989, is one of the most widely used models to describe consumers' intentions toward technical goods or services (42). It comprises four components: perceived utility, perceived ease of use, attitude, and intention to use. Perceived ease of use refers to the degree of ease associated with using a new product or service (42). On the other hand, perceived usefulness relates to how much individuals believe that novel goods or services can enhance productivity. Perceived usefulness typically has a positive impact on perceived ease of use, and together they influence customer behavior. Consumers' intentions have a direct impact on their actual behavior, as they affect the perceived benefits and overall attitude towards a product or service (42). While the TAM framework has received ongoing attention and validation, recent studies have examined the significance of the "attitude" variable as a mediator between variables and intention to use (43). However, the TAM framework has not been extensively explored in current research focusing on the acceptance of novel technology, specifically in developing countries (44,45,46,47). In Table 1, we provide an overview of several recent studies and their respective contexts.

**TABLE 1.** Summary of some of the related studies on ACC adoption.

| Authors | Sample Size | Study Location | Methodology | Important Findings |
|---|---|---|---|---|
| Mihale-Wilson et al. (48) | 278 | Germany | Choice-based conjoint analysis | Positive correlation between drivers' perceived usefulness of ACC and their WTP for it |
| Shin, J. et al. (49) | 675 | South Korea | Multiple discrete–continuous probit | There is considerable consumer preference heterogeneity for smart technology options, like ACC. |
| Bas, J. et al. (50) | 2853 | United States | discrete choice modeling | lower prices for the cruise system resulting from scale economies as market shares grow in time will translate in even higher adoption rates. |
| Larsson, A. F. (51) | 130 | Sweden | Questionnaire study | Using ACC might be better than a more perfect system, as it provides preparation for unexpected situations requiring the driver to reclaim control. |

| Hoedemaeker, M. et al. (31) | 38 | Netherland | The driving simulator | According to the findings, high-speed drivers tend to prefer the convenience offered by an ACC system, while low-speed drivers appreciate the system's practicality. |
| Sayer, J. et al. (52) | 66 | United States | Simulation | The results show that drivers who took part in the field test were no more likely to engage in secondary behaviors when driving with ACC and forward collision warning in comparison to manual control. |
| Asgari, H. et al. (53) | 1198 | United States | Structural equation modeling | People are more likely to pay for automated features when they believe that these features will provide them with benefits such as time and cost savings, convenience, stress reduction, and improved quality of life. |

The table provides a summary of different studies conducted in various countries, investigating the adoption and preferences related to ACC. The findings highlight factors influencing the WTP for ACC, including perceived usefulness, consumer preference heterogeneity, lower prices resulting from market growth, and the benefits associated with automated features. The studies also suggest that ACC offers convenience, practicality, and improved driving experiences, with drivers exhibiting similar behaviours and engagement levels compared to manual control. However, there is a significant gap in research regarding ACC adoption specifically in developing countries. This gap highlights the need to understand the unique context and challenges faced by developing countries in adopting ACC technology. Therefore, the present study aims to fill this gap by examining the factors influencing ACC adoption in a developing country context. By conducting research in these countries, this study aims to provide insights into the preferences, barriers, and potential benefits of ACC adoption in such regions. Understanding these factors is crucial for developing effective strategies to promote ACC adoption and addressing the specific challenges faced by developing countries in the realm of automotive technology.

## METHODS
### Questionnaire Design

In order to gather data for this study, a questionnaire was employed, consisting of two sections. The first section focused on capturing socio-economic characteristics, encompassing variables such as age, gender, education level, income, annual mileage, and prior usage of conventional cruise control. The second section was dedicated to exploring aspects related to ACC, with specific questions outlined in Table 2 for different respondent segments. These questions covered factors such as perceived ease of use, perceived usefulness, attitude towards use, WTP, and behavioral intention. Participants provided their responses on a Likert scale ranging from 1 (strongly disagree) to 5 (strongly agree), enabling a comprehensive assessment of their perspectives and opinions regarding ACC.

**TABLE 2.** Questions related to ACC which are used in Structrual Equation Modeling (SEM)

| Variable | Abbreviation | Question |
|---|---|---|
| Perceived ease of use | PEU01 | In my opinion, the car which has adaptive cruise control technology is suitable for my driving. |
| Perceived usefulness | PU01 | In my opinion, using a car with adaptive cruise control technology makes me more efficient while driving. |
|  | PU02 | In my opinion, using a car with adaptive cruise control technology may reduce traffic problems. |
|  | PU03 | In my opinion, using a car with adaptive cruise control technology reduces stress and improves driving. |
|  | PU04 | In my opinion, using a car with adaptive cruise control technology brings more safety for the passengers while driving. |
| Attitude towards using | ATU01 | I think it would be convenient for me to learn how to drive a car with adaptive cruise control technology. |
|  | ATU02 | In my opinion, a car with adaptive cruise control technology would be easy to drive. |
|  | ATU03 | I think it would be easy to achieve proficiency in driving a car with adaptive cruise control technology. |
| Behavioural intention | BI01 | If I have access to a car with adaptive cruise control technology, I will use it. |
|  | BI02 | If it is conceivable for me to buy a car with adaptive cruise control technology, I will buy it. |
|  | BI03 | I will add cars with adaptive cruise control to my car's dream list. |

**Data Collection**

In the pilot phase conducted in April 2022, a total of 50 paper-based questionnaires were collected. After analyzing the responses, certain questions were removed to streamline the survey's duration and improve the response rate. Subsequently, the main phase of the survey was carried out online, utilizing social media and website postings to distribute the survey link. Employing a snowball method, respondents were encouraged to share the survey link with their friends, relatives, and colleagues. The snowball approach is commonly employed in similar studies for data collection (54). Data collection for the main phase took place nationwide, spanning from June to November 2022, utilizing an online survey platform. Online survey platforms are widely recognized as reliable methods for collecting data (55). Ultimately, a total of 566 samples were collected. Excluding outlier data, a final sample of 453 observations was utilized for Modeling.

**Structural Equation and research hypotheses**

A structural equation, as shown in Figure 2, was used to investigate the adoption of ACC technology, in which 5 latent variables were used (1- perceived ease of use, 2- perceived usefulness, 3- attitude towards use, 4- willingness to pay, 5- behavioural intention). Drawing from existing literature, it has been assumed that the perceived ease of use has a positive effect on the perceived usefulness as well as the attitude towards the using, the perceived usefulness itself has a positive effect on the attitude towards the use and the attitude towards the use has a positive effect on the behavioural intention (56). This study further posits that WTP positively influences behavioral intention., so the following assumptions are suggested in this study:

H1: Perceived ease of use has a positive effect on perceived usefulness.

H2: Perceived ease of use has a positive effect on attitude towards using.

H3: Perceived usefulness has a positive effect on attitude towards using.

H4: Attitude towards using has a positive effect on behavioural intention.

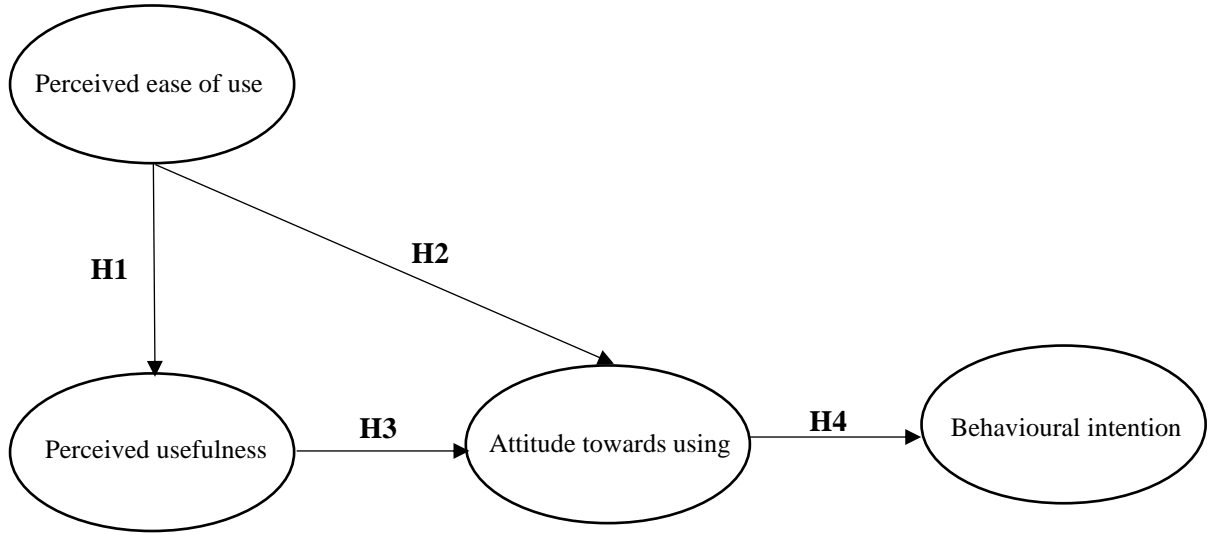

**Figure 2.** A structural equation to investigate the adoption of ACC technology

**Ordered logit**

Given that the dependent variable in this study is the payment amount for ACC, which is defined in a broad and sequential manner, a sequential model is employed. The hypothesis is based on choice models and involves the consideration of unobservable variables referred to as preference ratings, which indicate the extent to which individuals favor a particular option. Conversely, what is observed is a manifestation of the underlying construct, denoted as a latent variable. The models presented in this study take two general forms, namely logit and probit, as represented by equation 1.

$$Y_i = \beta X_i + \varepsilon_i \qquad (1$$

where, $Y_i$ is an unobservable dependent variable of the distance type related to observation i, which determines the priority of the desired option, $X_i$ is a vector of characteristics related to observation i (vector of independent variables), $\beta$ is a vector of model coefficients, and $\varepsilon_i$ is the random error part of observation i. It is assumed that $Z_i$ is the measurable reaction of the ordinal type related to observation i and the corresponding $Y_i$ and has M reaction groups $R_1, R_2, ..., R_M$ and resulting from the variable $Y_i$ in the form of equation 2. We also assume that there are M+1 real numbers $\mu_0, \mu_1, \mu_2, ..., \mu_M$ with the following conditions.

$$\mu_0 < \mu_1 < ... < \mu_M$$

$$Z_i \in R_j \leftrightarrow \mu_{j-1} \leq Y_i \leq \mu_j \qquad 1 \leq j \leq M \qquad (2$$

where $\mu_j$ is the upper threshold of reactive group j that determines $Z_i$ and its value. Since $Z_i$ is rank, it can be defined as a set of dummy variables according to equation 3 and 4.

$$Z_{ij} = \begin{cases} 1 & Z_i \in R_j \\ 0 & other \end{cases} \quad 1 \leq j \leq M \quad (3$$

$$Z_{ij} = 1 \leftrightarrow \mu_{j-1} \leq Y_i \leq \mu_j \leftrightarrow \mu_{j-1} \leq \sum_k \beta_k X_{ki} + \varepsilon_i \leq \mu_j \leftrightarrow$$

$$\frac{(\mu_{j-1} - \sum_k \beta_k X_{ki} + \varepsilon_i)}{\sigma} \leq \frac{\varepsilon_i}{\sigma} \leq \frac{(\mu_j - \sum_k \beta_k X_{ki} + \varepsilon_i)}{\sigma} \quad (4$$

Where k=0, 1, ..., k indicates k is the feature of the option and $\sigma$ is the standard deviation of $\varepsilon_i$, and if $\varepsilon_i$ has a Gumbel distribution, it is an ordinal logit model. In this study, the ordinal logit model is used for the WTP for adaptive cruise control.

The dependent variable in this study comprises seven options, representing different levels of WTP. These options include: (1) not willing to pay any amount, (2) less than 50 million IRR, (3) between 50 and 150 million IRR, (4) between 150 and 300 million IRR, (5) between 300 and 500 million IRR, (6) between 500 and 800 million IRR, and (7) above 800 million IRR. The variables included in the model consist of car price exceeding 5 billion IRR, women who have cruise control, perceived ease of use, and efficiency, which are the outcomes of the structural equation. Given the seven options for the dependent variable, the model yields six thresholds for option selection. If the obtained value from the model is lower than $\mu_0$, option 1 is chosen, and if it is higher than $\mu_5$, option 7 is selected.

**RESULTS**

**Descriptive analysis**

A descriptive analysis was conducted to examine the data collected in this study. The results are presented in Table 3. The analysis revealed that approximately 73.3% of the respondents were male. Furthermore, around 45.3% of the participants fell within the age range of 20 to 29 years. In terms of education, 46.6% of the respondents were either undergraduate or graduate students. It was found that 47.9% of the participants had an income of less than 50 million IRR, while 25.2% had an income ranging from 50 to 100 million IRR. When considering driving experience, 17.9% of the respondents held a driving license for less than one year, and 40.6% had prior experience using CC. Additionally, 24.7% of the respondents reported owning a car equipped with CC technology. The predominance of male respondents suggests that men tend to drive more frequently in developing countries (57). Given that younger individuals are more active on social media, it is not surprising that the statistical sample predominantly consisted of individuals between the ages of 20 and 29. The higher representation of respondents with bachelor's and master's degrees reflects the educational background of the sample population. The majority of respondents had an income below 50 million IRR. The descriptive statistics of the participants are presented in Table 3.

**TABLE 3.** Variables with their categories and descriptions

| Variable | Description | Number | Percentage |
|---|---|---|---|
| Gender | Female | 119 | 26.3% |
|  | Male | 334 | 73.7% |
| Age | 18 - 19 | 70 | 15.5% |
|  | 20 - 29 | 205 | 45.3% |
|  | 30 - 39 | 72 | 15.9% |
|  | 40 - 49 | 52 | 11.5% |
|  | 50 - 59 | 48 | 10.6% |
|  | Over 60 | 6 | 1.3% |
| Education | High school | 2 | 0.4% |
|  | diploma | 80 | 17.7% |
|  | Associate degree | 27 | 6.0% |
|  | Undergraduate or graduate student | 211 | 46.6% |
|  | Master's student or graduate | 89 | 19.6% |
|  | PhD student or PhD graduate | 44 | 9.7% |
| Income/ month | Less than 50 million IRR | 217 | 47.9% |
|  | 50 to 100 million IRR | 114 | 25.2% |
|  | 100 to 150 million IRR | 64 | 14.1% |
|  | 150 to 200 million IRR | 35 | 7.7% |
|  | 200 to 500 million IRR | 17 | 3.8% |
|  | Over 500 million IRR | 6 | 1.3% |
| Years of having a driving license | Less than 1 year | 81 | 17.9% |
|  | Between 1 to 2 years | 72 | 15.9% |
|  | Between 2 to 4 years | 65 | 14.3% |
|  | Between 5 to 9 years | 79 | 17.4% |
|  | Between 10 to 19 years | 91 | 20.1% |
|  | Over 20 years | 65 | 14.3% |
| Having experience using CC | Yes | 269 | 59.4% |
|  | No | 184 | 40.6% |
| Having a car with CC technology | Yes | 112 | 24.7% |
|  | No | 341 | 75.3% |

**Structural Equation Modelling for TAM of ACC**

SEM was performed using SmartPLS software, and the results are presented in Figure 3. The reliability of the obtained measures was assessed using Cronbach's alpha, with all values exceeding 0.7 except for WTP, which had a Cronbach's alpha of 0.633. However, this lower value can be disregarded considering the composite reliability. Table 4 demonstrates that all variables exhibit high reliability based on composite reliability scores.

**TABLE 4.** Reliability tests

| Variables | Cronbach's Alpha | Composite Reliability | Composite Reliability |
|---|---|---|---|
| Attitude Towards Using | 0.862 | 0.871 | 0.916 |
| Behavioural Intention | 0.802 | 0.809 | 0.883 |
| Perceived usefulness | 0.777 | 0.796 | 0.856 |

In Figure 2, the coefficients depicted on the arrows between variables represent the path coefficients indicating the degree of influence of each variable on the dependent variable. The coefficients on the arrows of the indicators for each variable represent the factor loadings of the corresponding indicators. Notably, all factor loadings for the aforementioned variables are acceptable at 0.4. Table 5 presents the values of R2 and adjusted R2.

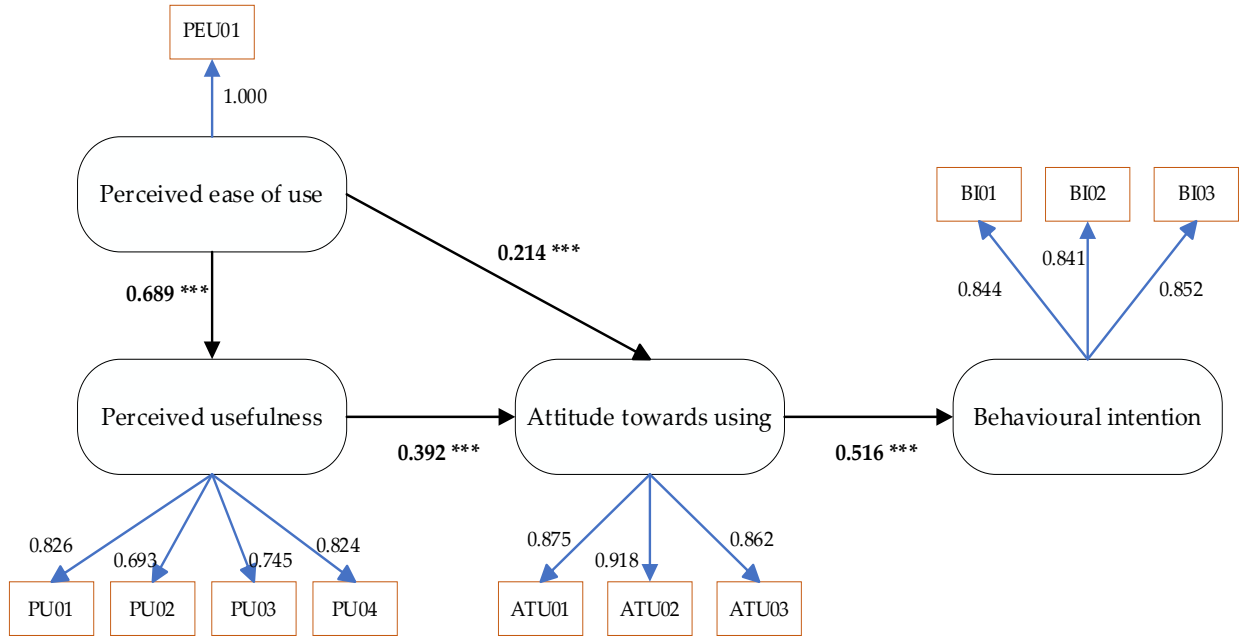

Note: p* < 0.01; p** < 0.05; p*** < 0.001

**Figure 3.** The TAM for adaptive cruise control

**TABLE 5.** Values of $R^2$ and adjusted $R^2$

| Variables | $R^2$ | Adjusted $R^2$ |
|---|---|---|
| Attitude Towards Using | 0.315 | 0.312 |
| Behavioural Intention | 0.266 | 0.265 |
| Perceived usefulness | 0.475 | 0.474 |

The VIF tests for the variables are given in Table 6. As they are below 10 they are acceptable and no multi-collinearity can be seen in this model. Also, Figure 4 indicates the correlation matrix between indicators. In SEM, the indicators of a latent variable are expected to be correlated with each other, as they share a common source of variance which is the latent variable. If indicators are not correlated with each other, it may indicate that the latent variable is not suitable for explaining the observed data. Having a reasonable amount of correlation among indicators is desirable, but very high correlations may suggest that some indicators are redundant and could be measured more efficiently with a smaller set of indicators (58).

**TABLE 6.** The VIF tests for the variables

| Variable | ATU01 | ATU02 | ATU03 | BI01 | BI02 | BI03 | PE01 | PE02 | PE03 | PE04 | PEU01 |
|---|---|---|---|---|---|---|---|---|---|---|---|
| VIF | 2.094 | 2.633 | 2.124 | 1.662 | 1.855 | 1.693 | 1.707 | 1.406 | 1.490 | 1.738 | 1.000 |

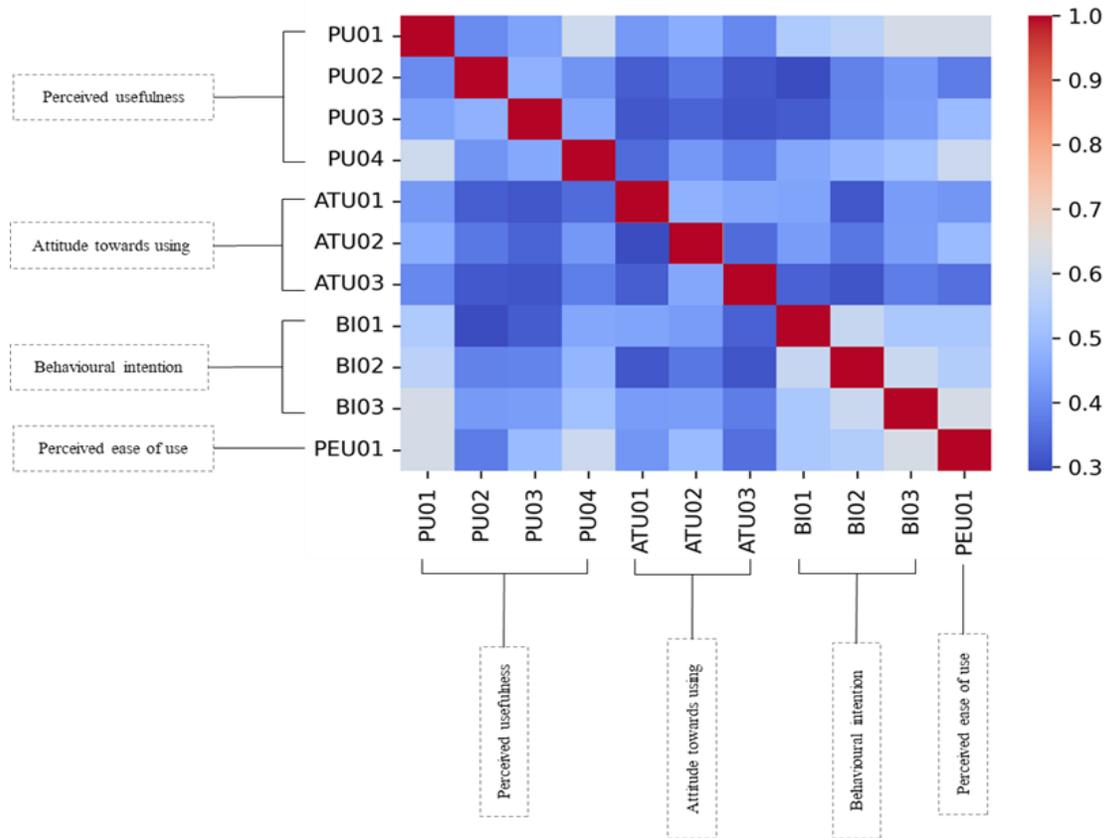

**Figure 4.** Correlation matrix of indicators

Table 7 presents the results of the t-tests and corresponding p-values for each relationship examined. The statistical analysis indicates that all the defined relationships are statistically significant, based on their significance levels and t-test results. Furthermore, all the assumptions proposed in the methodology section of this study have been supported. Specifically, ease of use demonstrates a significant direct effect, accounting for 68.9% of the variance in perceived usefulness and 21.4% in attitude towards using. Perceived usefulness, in turn, exerts a significant direct effect of 39.2% on attitude towards using. Lastly, attitude towards using demonstrates a substantial direct effect of 51.6% on behavioral intention.

**TABLE 7.** Statistical analyzing of hypothesis in TAM of ACC

| Relationships | Original sample | Sample mean | Standard deviation | T value | P value |
|---|---|---|---|---|---|
| Attitude Toward_Using To Behavioral_Intention | 0.516 | 0.518 | 0.048 | 10.722 | 0.000*** |
| Perceived Ease_of Use To Attitude Toward_Using | 0.214 | 0.212 | 0.062 | 3.440 | 0.001*** |
| Perceived Ease_of Use To Perceived_Usefulness | 0.689 | 0.690 | 0.030 | 22.717 | 0.000*** |
| Perceived_Usefulness To Attitude Toward_Using | 0.392 | 0.393 | 0.067 | 5.849 | 0.000*** |

Note: p* < 0.01; p** < 0.05; p*** < 0.001

The analysis presented in Table 8 employed an ordered logit model to examine the relationship between car price, perceived ease of use, perceived usefulness, and the ownership of cars with conventional

CC among women. The findings indicate that individuals owning cars with a price exceeding 5 billion rials exhibit a higher inclination towards adoption compared to others. Within the framework of the TAM, the latent variables of perceived ease of use and perceived usefulness demonstrate stronger associations with the WTP model. While individuals across genders aspire to have a more convenient driving experience, women exhibit a higher level of interest compared to men. This model further reveals that women who have prior experience with CC are significantly more eager to purchase ACC for their cars than men. This may be attributed to the preference of men for challenging and riskier driving experiences, which ACC does not provide. Conversely, many women are less inclined towards high-speed driving, making them more receptive to the adoption of ACC. The p-values associated with car price, women with CC experience, and perceived ease of use are satisfactory and deemed acceptable. All four parameters positively contribute to the utility function of WTP, with car price exerting the strongest impact and perceived usefulness demonstrating the lowest effect.

**TABLE 8.** The result of ordered logit model for WTP of ACC

| Variables | Coefficient | Standard error | T value | P value |
|---|---|---|---|---|
| 1.PCAR2 | 0.7388865 | 0.2649412 | 2.79 | 0.005** |
| Perceived Ease_of Use | 0.4126363 | 0.1305018 | 3.16 | 0.002** |
| Perceived_Usefulness | 0.2587127 | 0.136767 | 1.89 | 0.059* |
| 1.FMCC | 0.6894204 | 0.3347697 | 2.06 | 0.039** |
| **Cut-offs** | | | | |
| $\mu_0$ | -1.266797 | 0.1117171 | | |
| $\mu_1$ | 0.8923196 | 0.1238362 | | |
| $\mu_2$ | 2.458778 | 0.1858625 | | |
| $\mu_3$ | 3.386791 | 0.2418896 | | |
| $\mu_4$ | 4.065735 | 0.3211643 | | |
| $\mu_5$ | 5.098537 | 0.508155 | | |
| LR chi2(4) = 56.42 | | | | |
| Prob > chi2 = 0.0000 | | | | |
| Log likelihood = -603.01146 | | | | |
| Pseudo R2 = 0.0447 | | | | |

Note: p* < 0.01; p** < 0.05; p*** < 0.001

A dummy variable named PCAR2 has been assigned a value of 1 for cars with a price exceeding 5 billion rials in the table. The variables Perceived Ease of Use and Perceived Usefulness are latent variables that have been utilized in the TAM of ACC. Furthermore, a dummy variable labeled FMCC represents female car owners who possess cars equipped with cruise control.

The cut-off values in the model represent the thresholds that determine different levels of WTP for ACC. Each specific cut-off value signifies the point at which respondents are more likely to choose a higher category of WTP. The log-likelihood value serves as a valuable indicator of the model's predictive capability. With a lower log-likelihood value (-603.01146), the model exhibits a better fit, indicating its ability to accurately capture the underlying patterns and relationships within the data. In addition, the Pseudo R2, which measures the proportion of variance explained by the model, provides further insight into its overall performance. The analysis of the model's goodness of fit reveals its statistical significance, indicating an alignment between the model and the observed data. Figure 5 shows the marginal effect of independent variables in the model. This figure expresses the amount of change in the probability of the options based on the change in each unit of the variables.

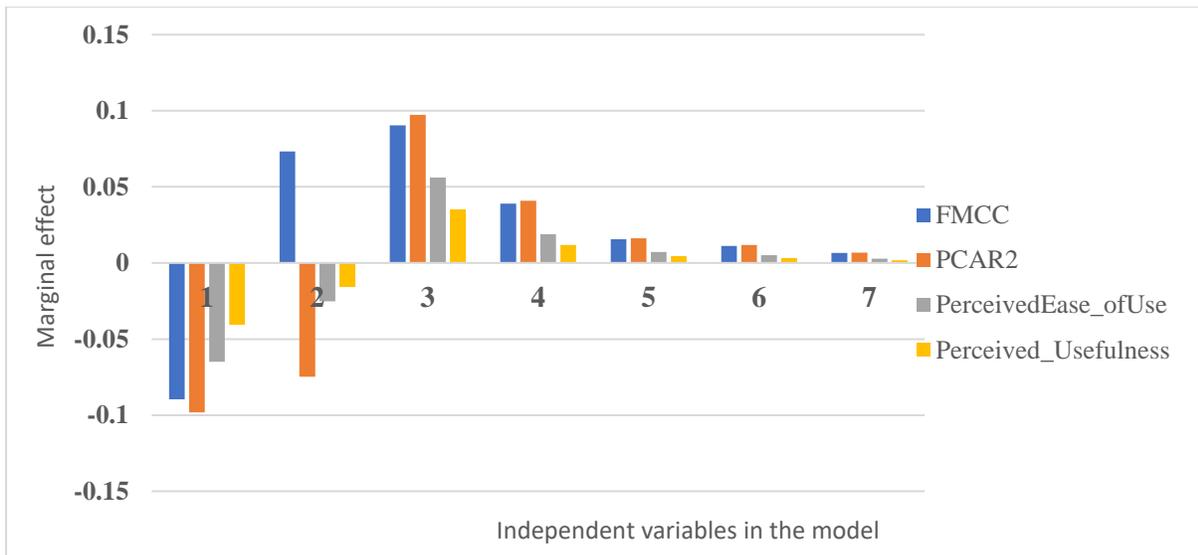

**Figure 5.** The marginal effect of variables on willingness to pay for ACC

## DISCUSSION

The adoption of ACC is influenced by various factors, including perceived ease of use, perceived usefulness, attitude towards using the technology, and behavioral intention. In this study, all of these factors were found to be positively accepted by participants, despite the low prevalence of ACC in Iran. This suggests that Iranian individuals are receptive to new technologies, regardless of their prior knowledge or exposure. However, when participants were asked about their WTP for the installation of ACC, their responses were less favourable. The difference in the value of the US dollar and the Iranian Rial contributed to a lack of enthusiasm for paying the high costs associated with ACC installation, with only the capitalist class showing interest.

Considering the environmental impact, both CC and ACC technologies have the potential to significantly reduce fuel consumption, thereby directly contributing to the reduction of air pollution and global warming. Therefore, policymakers in the car production and transportation management sectors are urged to consider strategies that promote environmentally friendly practices. One approach could involve equipping vehicles priced at 5 billion Rials and above with ACC, while installing CC on vehicles priced below 5 billion Rials. This would represent a substantial step towards environmental conservation. Alternatively, implementing subsidies or offering incentives such as reduced toll fees or parking costs for vehicles equipped with ACC and CC could significantly increase their adoption rates, leading to improved air quality and a more sustainable future.

## CONCLUSION

This study focused on understanding the factors influencing the adoption of ACC technology among drivers with different characteristics in Iran. ACC technology revolutionizes the driving experience by offering a multitude of advantages that enhance safety, improve traffic flow, and promote environmental sustainability. ACC utilizes advanced sensors and intelligent algorithms to maintain a safe distance from the vehicle ahead, reducing the risk of rear-end collisions caused by human error. It ensures a consistent and safe following distance, mitigating abrupt braking or acceleration. Moreover, ACC reduces driver stress and fatigue by automatically adjusting speed and distance in congested traffic or during long journeys, allowing drivers to focus more on the road ahead. In addition to safety and comfort, ACC plays a pivotal role in optimizing traffic efficiency. By maintaining a steady speed and following distance, it minimizes unnecessary braking and acceleration, resulting in smoother traffic flow and reduced congestion (59). This not only shortens commute times but also improves fuel efficiency and reduces emissions, leading to a more sustainable and eco-friendly transportation system. ACC's collision avoidance features, such as collision warning and automatic emergency braking, add an extra layer of protection, alerting drivers to potential hazards and assisting in preventing accidents. The findings shed light on the target audience group and strategies for penetration rate. (60)

Through the analysis of the SEM for the TAM of ACC, it was evident that people from various demographic backgrounds, including gender, educational qualification, income, and car price, exhibited a strong desire to have or install ACC on their vehicles. This suggests a widespread acceptance and positive attitude towards the technology among drivers in Iran. These findings are in line with the diffusion theory's concept of early majority and late majority adopters, who tend to follow the trends set by innovators and early adopters. Notably, the results from the ordered logit model exploring the WTP for ACC installation revealed some intriguing insights. Although individuals who initially expressed agreement with ACC adoption did not demonstrate a strong tendency to pay a substantial amount for the technology, it was observed that women, who prioritize driving comfort, exhibited a greater inclination to pay for ACC installation. In addition, capitalists, known for their tendency to adopt up-to-date technologies, displayed eagerness to have ACC.

Considering these findings, it is recommended to target early adopters, such as women and capitalists, who have shown a higher propensity towards ACC adoption. Their enthusiasm and positive experiences can serve as influential factors in promoting ACC among their peers and the broader population. To increase penetration rate, it is crucial to emphasize the benefits of ACC, particularly its ability to enhance traffic safety and contribute to a more sustainable environment. Policy-making efforts should focus on educating potential adopters about the user-friendly features of ACC, addressing any concerns or misconceptions they may have. Starting with targeted campaigns and awareness programs tailored towards the identified audience groups, the adoption of ACC can gain momentum. Collaborations with automotive manufacturers and dealerships to integrate ACC as a standard or optional feature in new car models can also contribute to high penetration rate. Additionally, providing accessible information and support for ACC installation and usage will help potential adopters make informed decisions.

While this research offers valuable insights, it is important to acknowledge its limitations. The online distribution of the questionnaire may have introduced a bias by excluding individuals without internet access. Furthermore, the reliance on a brief film to gauge respondents' understanding of ACC due to its low prevalence in Iran could have influenced their responses. The significant disparity between the US dollar and Iranian rial complicated the formulation of WTP questions, potentially limiting the applicability of the findings. Moving forward, it is recommended that future investigations concentrate on assessing the necessary infrastructure for ACC integration in Iran, identifying areas for improvement. In addition, exploring public perceptions and attitudes towards autonomous vehicles can provide valuable insights, contributing to a comprehensive understanding of factors influencing ACC adoption. By incorporating public perspectives, decision-making processes can be better informed to facilitate the implementation of ACC technology in Iran, aligning with the needs and expectations of the population.

**ACKNOWLEDGENT**